\documentclass{llncs}
\usepackage{amsmath,amsfonts,amssymb}
\usepackage{graphicx}
\usepackage{nicefrac}

\newcommand{\eg}{e.\,g.~}
\newcommand{\ie}{i.\,e.~}

\renewcommand{\vec}[1]{\boldsymbol{#1}}
\newcommand{\mat}[1]{\boldsymbol{#1}}

\begin{document}

\title{DeepDRR -- A Catalyst for Machine Learning in Fluoroscopy-guided Procedures }
\titlerunning{DeepDRR}  
%
\author{Mathias~Unberath\inst{1,}\thanks{Both authors contributed equally.} \and Jan-Nico~Zaech\inst{1,2,\star} \and Sing~Chun~Lee\inst{1} \and Bastian~Bier\inst{1,2} \and Javad~Fotouhi\inst{1} \and Mehran~Armand\inst{4} \and Nassir~Navab\inst{3}}
\authorrunning{Unberath and Z{\"a}ch et al.} 
\institute{Computer Aided Medical Procedures, Johns Hopkins University\\
\and
Pattern Recognition Lab, Friedrich-Alexander-Universit{\"a}t Erlangen-N{\"u}rnberg
\and
Applied Physics Laboratory, Johns Hopkins University
}

\maketitle              

\begin{abstract}
Machine learning-based approaches outperform competing methods in most disciplines relevant to diagnostic radiology. Interventional radiology, however, has not yet benefited substantially from the advent of deep learning, in particular because of two reasons: 1) Most images acquired during the procedure are never archived and are thus not available for learning, and 2) even if they were available, annotations would be a severe challenge due to the vast amounts of data. When considering fluoroscopy-guided procedures, an interesting alternative to true interventional fluoroscopy is \emph{in silico} simulation of the procedure from 3D diagnostic CT. In this case, labeling is comparably easy and potentially readily available, yet, the appropriateness of resulting synthetic data is dependent on the forward model. In this work, we propose DeepDRR, a framework for fast and realistic simulation of fluoroscopy and digital radiography from CT scans, tightly integrated with the software platforms native to deep learning. We use machine learning for material decomposition and scatter estimation in 3D and 2D, respectively, combined with analytic forward projection and noise injection to achieve the required performance. On the example of anatomical landmark detection in X-ray images of the pelvis, we demonstrate that machine learning models trained on DeepDRRs generalize to unseen clinically acquired data without the need for re-training or domain adaptation. Our results are promising and promote the establishment of machine learning in fluoroscopy-guided procedures.
\end{abstract}
\section{Introduction}
The advent of convolutional neural networks (ConvNets) for classification, regression, and prediction tasks, currently most commonly referred to as deep learning, has brought substantial improvements to many well studied problems in computer vision, and more recently, medical image computing. This field is dominated by diagnostic imaging tasks where 1) all image data are archived, 2) learning targets, in particular annotations of any kind, exist traditionally~\cite{kooi2017large} or can be approximated~\cite{roy2017error}, and 3) comparably simple augmentation strategies, such as rigid and non-rigid displacements~\cite{milletari2016v}, ease the limited data problem.\\
Unfortunately, the situation is more complicated in interventional imaging, particularly in 2D fluoroscopy-guided procedures. First, while many X-ray images are acquired for procedural guidance, only very few radiographs that document the procedural outcome are archived suggesting a severe lack of meaningful data. Second, learning targets are not well established or defined; and third, there is great variability in the data, \eg due to different surgical tools present in the images, which challenges meaningful augmentation. Consequently, substantial amounts of clinical data must be collected and annotated to enable machine learning for fluoroscopy-guided procedures. Despite clear opportunities, in particular for prediction tasks, very little work has considered learning in this context~\cite{li2016automatic,terunuma2017novel,ambrosini2017fully,ma2017fast}.\\
A promising approach to tackling the above challenges is \emph{in silico} fluoroscopy generation from diagnostic 3D CT, most commonly referred to as digitally reconstructed radiographs (DRRs)~\cite{li2016automatic,terunuma2017novel}. Rendering DRRs from CT provides fluoroscopy in known geometry, but more importantly: Annotation and augmentation can be performed on the 3D CT substantially reducing the workload and promoting valid image characteristics, respectively. However, machine learning models trained on DRRs do not generalize to clinical data since traditional DRR generation, \eg as in~\cite{russakoff2005fast,li2016automatic}, does not accurately model X-ray image formation. To overcome this limitation we propose DeepDRR, an easy-to-use framework for realistic DRR generation from CT volumes targeted at the machine learning community. On the example of view independent anatomical landmark detection in pelvic trauma surgery~\cite{landmarksMICCAI}, we demonstrate that training on DeepDRRs enables direct application of the learned model to clinical data without the need for re-training or domain adaptation.

\section{Methods}

\subsection{Background and Requirements} 
DRR generation considers the problem of finding detector responses given a particular imaging geometry according to Beer-Lambert law~\cite{hubbell1995tables}. 
Methods for \emph{in silico} generation of DRRs can be grouped in analytic and statistical approaches, \ie ray-tracing and Monte Carlo (MC) simulation, respectively. Ray-tracing algorithms are computationally efficient since the attenuated photon fluence of a detector pixel is determined by computing total attenuation along a 3D line that then applies to all photons emitted in that direction~\cite{russakoff2005fast}. Commonly, ray-tracing only considers a single material in the mono-energetic case and thus fails to model beam hardening. In addition and since ray-tracing is analytic, statistical processes during image formation, such as scattering, cannot be modeled. Conversely, MC methods simulate single photon transport by evaluating the probability of photon-matter interaction, the sequence of which determines attenuation~\cite{badal_accelerating_2009}. Since the probability of interaction is inherently material and energy dependent, MC simulations require material decomposition in CT that is usually achieved by thresholding of CT values (Houndfield units, HU)~\cite{schneider2000correlation} and spectra of the emitter~\cite{badal_accelerating_2009}. As a consequence, MC is very realistic. Unfortunately, for training-set-size DRR generation on conventional hardware, MC is prohibitively expensive. As an example, accelerated MC simulation~\cite{badal_accelerating_2009} on an NVIDIA Titan Xp takes $\approx4$\,hours for a single X-ray image with $10^{10}$\,photons. 
To leverage the advantages of MC simulations in clinical practice, the medical physics community provides further acceleration strategies if prior knowledge on the problem exists. A well studied example is variance reduction for scatter correction in cone-beam CT, since scatter is of low frequency~\cite{sisniega2013monte}.\\
Unfortunately, several challenges remain that hinder the implementation of realistic \emph{in silico} X-ray generation for machine learning applications. We have identified the following fundamental challenges at the interface of machine learning and medical physics that must be overcome to establish realistic simulation in the machine learning community: 1) Tools designed for machine learning must seamlessly integrate with the common frameworks. 2) Training requires many images so data generation must be fast and automatic. 3) Simulation must be realistic: Both analytic and statistic processes such as beam-hardening and scatter, respectively, must be modeled.

\begin{figure}[tb] 
\centering
{\includegraphics[width=1.0\textwidth]{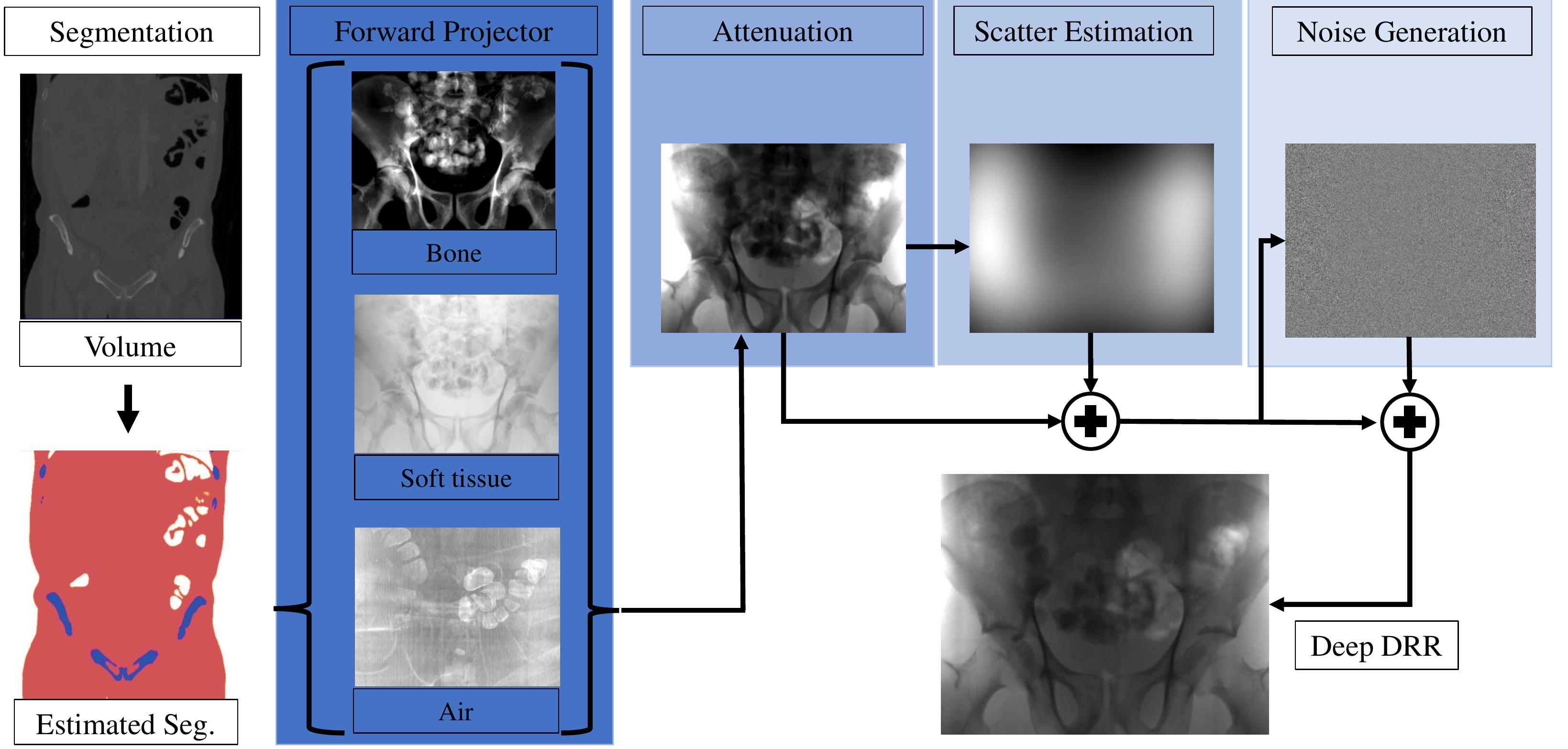}} 
\caption{Schematic overview of DeepDRR.} 
\label{fig:overview}
\end{figure}

\subsection{DeepDRR}
\label{sec:deepDRR}

\paragraph{Overview:} We propose DeepDRR, a Python, PyCUDA, and PyTorch-based framework for fast and automatic simulation of X-ray images from CT data. It consists of 4 major modules: 1) Material decomposition in CT volumes using a deep segmentation ConvNet; 2) A material- and spectrum-aware ray-tracing forward projector; 3) A neural network-based Rayleigh scatter estimation; and 4) Quantum and electronic readout noise injection. The individual steps of DeepDRR are visualized schematically in Fig.~\ref{fig:overview} and explained in greater detail in the remainder of this section. The fully automated pipeline will be made open source upon publication of this manuscript.

\paragraph{Material Decomposition:} Material decomposition in 3D CT for MC simulation is traditionally accomplished by thresholding, since a given material has a characteristic HU range~\cite{schneider2000correlation}. This works well for large HU discrepancies, \eg air ($[-1000]\,$HU) and bone ($[200,3000]\,$HU), but may fail otherwise, particularly between soft tissue ($[-150,300]\,$HU) and bone in presence of low mineral density. This is problematic since, despite similar HU, the attenuation characteristic of bone is substantially different of soft tissue~\cite{hubbell1995tables}. Within this work, we use a deep volumetric ConvNet adapted from~\cite{milletari2016v} to automatically decompose air, soft tissue, and bone in CT volumes. The ConvNet is of encoder-decoder structure with skip-ahead connections to retain information of high spatial resolution while enabling large receptive fields. The ConvNet is trained on patches with $128\times128\times128$\,voxels with voxel sizes of $0.86\times0.86\times1.0$\,mm yielding a material map $M(\vec{x})$ that assigns a candidate material to each 3D point $\vec{x}$. 
We used the multi-class Dice loss as the optimization target. 12 whole-body CT data were manually annotated, and then split: 10 for training, and 2 for validation and testing. 
Training was performed over 600 epochs until convergence where, in each epoch, one patch from every volume was randomly extracted. 
During application, patches of $128\times128\times128$\,voxels are fed-forward with stride of 64 since only labels for the central $64\times64\times64$\,voxels are accepted.

\begin{figure}[tb] 
\centering
{\includegraphics[width=0.9\textwidth]{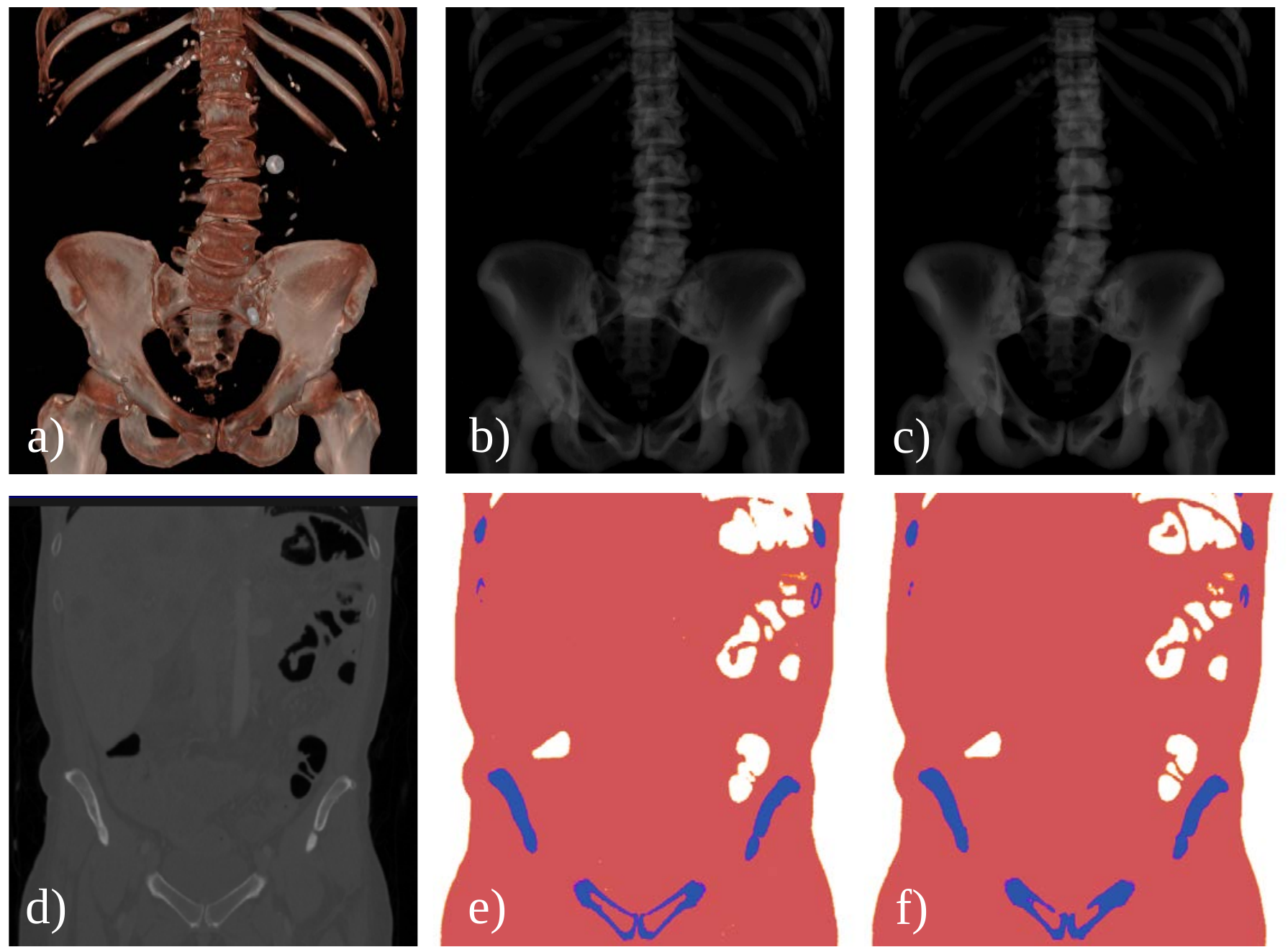}} 
\caption{Representative results of the segmentation ConvNets. From left to right, the columns show input volume, manual segmentation, and ConvNet result. The top rows shows volume renderings of the bony anatomy and respective label, while the bottom row shows a coronal slice through the volumes.} 
\label{fig:segmentationResults}
\end{figure}

\paragraph{Analytic Primary Computation:} Once segmentations of the considered materials $M=\{\text{air, soft tissue, bone}\}$ are available, the contribution of each material to the total attenuation density at detector position $\vec{u}$ are computed using a given geometry (defined by projection matrix $\mat{P}\in\mathbb{R}^{3\times4}$) and X-ray spectral density $p_0(E)$ via ray-tracing:
\begin{align}
p(\vec{u}) &= \int p(E,\vec{u})\mathrm{d}E\nonumber\\
&= \int p_0(E)\exp\left( \sum_{m\in M} \delta\left(m,M(\vec{x})\right)  (\nicefrac{\mu}{\rho})_m(E) \int \rho(\vec{x})\, \mathrm{d}\vec{l}_{\vec{u}} \right)\mathrm{d}E\,,
\end{align}
where $\delta\left(\cdot,\cdot\right)$ is the Kronecker delta, $\vec{l}_{\vec{u}}$ is the 3D ray connecting the source position and 3D location of detector pixel $\vec{u}$ determined by $\mat{P}$, $(\nicefrac{\mu}{\rho})_m(E)$ is the material and energy dependent linear attenuation coefficient~\cite{hubbell1995tables}, and $\rho(\vec{x})$ is the material density at position $\vec{x}$ derived from HU values. The projection domain image $p(\vec{u})$ is then used as input to our scatter prediction ConvNet. 

\paragraph{Learning-based Scatter Estimation:}
Traditional scatter estimation relies on variance-reduced MC simulations~\cite{sisniega2013monte}, which requires a complete MC setup. Recent approaches to scatter estimation via ConvNets outperform kernel based methods~\cite{scatter2018SPIE} while retaining the low computational demand. In addition, they inherently integrate with deep learning software environments. We define a ten layer ConvNet, where the first six layers generate Rayleigh scatter estimates and the last four layers, with $31 \times31$ kernels and a single channel, ensure smoothness. 
The network was trained on 330 images generated via MC simulation~\cite{badal_accelerating_2009}, augmented by random rotations and reflections. The last three layers where trained using pre-training of the preceding layers. The input to the network is downsampled to $128\times128$\,pixels.

\paragraph{Noise Injection:} After adding scatter, $p(\vec{u})$ expresses the energy deposited by a photon in detector pixel $\vec{u}$. The number of photons is estimated as:
\begin{equation}
\label{eq:label}
N(\vec{u}) = \sum_{E} \frac{p(E,\vec{u})}{E} N_0\,,
\end{equation}
to obtain the number of registered photons $N(\vec{u})$ and perform realistic noise injection. In Eq.~\ref{eq:label}, $N_0$ (potentially location dependent $N_0(\vec{u})$, \eg due to bow-tie filters) is the emitted number of photons per pixel. Noise in X-ray images is a composite of uncorrelated quantum noise due to photon statistics that becomes correlated due to pixel crosstalk, and correlated readout noise~\cite{zhang2014noise}. Due to beam hardening, the spectrum arriving at any detector pixel differs. To account for this fact in the Poisson noise model, we compute a mean photon energy for each pixel by $\bar{E}(\vec{u})$ and estimate quantum noise as $\nicefrac{\bar{E}}{N_0}(p_{Poisson}(N)-N)$, where $p_{Poisson}$ is the Poisson generating function. Since real flat panel detectors suffer from pixel crosstalk, we correlate the quantum noise of neighboring pixels by convolving the noise signal with a blurring kernel~\cite{zhang2014noise}.
The second major noise component is electronic readout noise. Electronic noise is signal independent and can be modeled as additive Gaussian noise with correlation along rows  due to sequential readout~\cite{zhang2014noise}. Finally, we obtain a realistically simulated DRR.

\section{Experiments and Results}

\subsection{Framework Validation}
Since forward projection and noise injection are analytic processes, we only assess the prediction accuracy of the proposed ConvNets for volumetric segmentation and projection domain scatter estimation.\\
For volumetric segmentation of air, soft tissue, and bone in CT volumes, we found a misclassification rate of $(2.03\pm 3.63)$\,\% which is in line with results reported in previous studies using this architecture~\cite{milletari2016v}. Representative results on the testing set are shown in Fig.~\ref{fig:segmentationResults}. For scatter estimation, the evaluation on a test set consisting of 30 image yields a mean squared error of $6.4\,$\% of the total scatter image energy.\\
Simulation of one X-ray projection image with $615\times479$\,pixels took $2.0$\,s irrespective of number of photons used.

\begin{figure}[tb] 
\centering
{\includegraphics[width=1.0\textwidth]{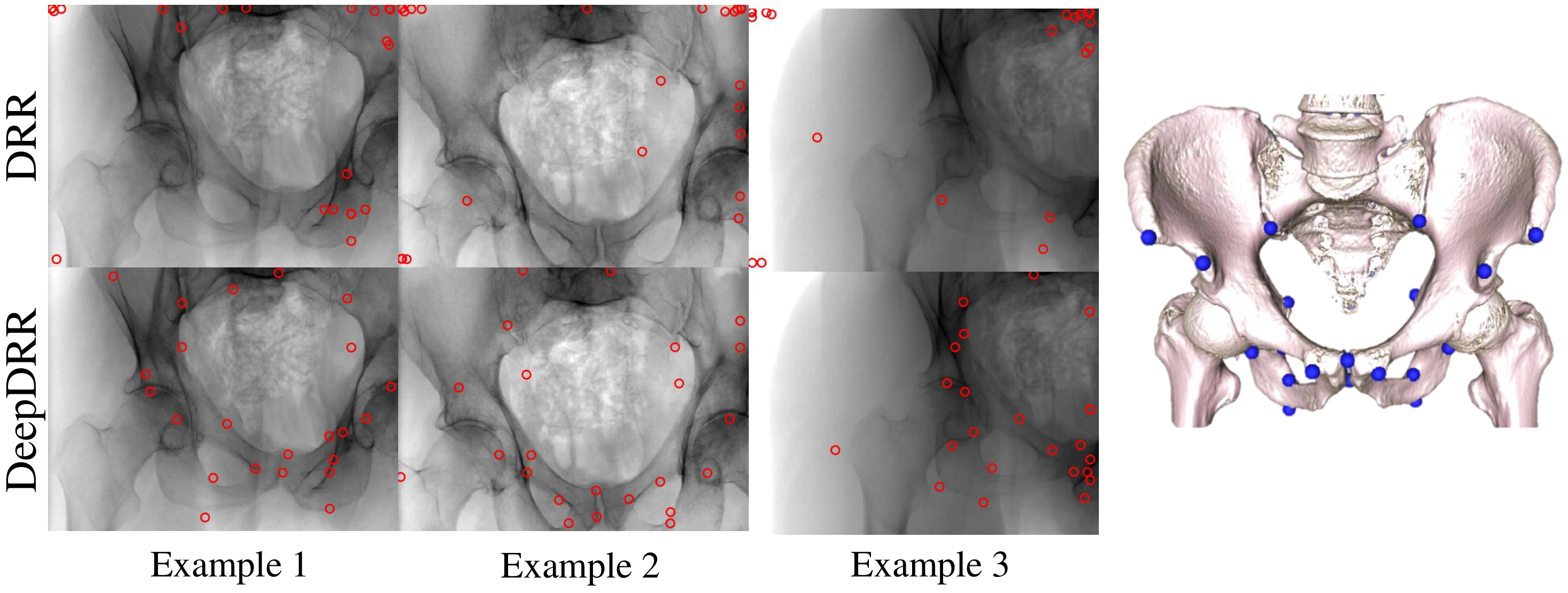}} 
\caption{Anatomical landmark detection on real data of cadaveric specimen using the method detailed in~\cite{landmarksMICCAI}. Top row: Detection results of a model trained on conventional DRRs. Bottom row: Detections of a model trained on the proposed DeepDRRs. No domain adaption or re-training was performed. Right-most image: Schematic illustration of desired landmark locations shown on a training set sample.} 
\label{fig:landmarkBelief}
\end{figure} 

\subsection{Task-based Evaluation}
Fundamentally, the goal of DeepDRR is to enable the learning of models on synthetically generated data that generalizes to unseen clinical fluoroscopy without re-training or other domain adaptation strategies. To this end, we consider anatomical landmark detection in X-ray images of the pelvis from arbitrary views~\cite{landmarksMICCAI}. The authors annotated 23 anatomical landmarks in CT volumes of the pelvis (Fig.~\ref{fig:landmarkBelief}, last column) and generated DRRs with annotations on a spherical segment covering $120^\circ$ and $90^\circ$ in RAO/LAO and CRAN/CAUD, respectively. Then, a sequential prediction framework is learned and, upon convergence, used to predict the 23 anatomical landmarks in unseen, real X-ray images of cadaver studies. The network is learned twice: First, on conventionally generated DRRs assuming a single material and mono-energetic spectrum, and second, on DeepDRRs as described in Sec.~\ref{sec:deepDRR}. Images had $615\times479$\,pixels with $0.616^2\,$mm pixel size. We used the spectrum of a tungsten anode operated at $120\,$kV with $4.3\,$mm aluminum and assumed a high-dose acquisition with $5\cdot10^{5}$\,photons per pixel.\\
In Fig.~\ref{fig:landmarkBelief} we show representative detections of the sequential prediction framework on unseen, clinical data acquired using a flat panel C-arm system (Siemens Cios Fusion, Siemens Healthcare GmbH, Germany) during cadaver studies. As expected, the model trained on conventional DRRs (upper row) fails to predict anatomical landmark locations on clinical data, while the model trained on DeepDRRs produces accurate predictions even on partial anatomy. In addition, we would like to refer to the comprehensive results reported in~\cite{landmarksMICCAI} that were achieved using training on the proposed DeepDRRs. 

\section{Discussion and Conclusion}
We proposed DeepDRR, a framework for fast and realistic generation of synthetic X-ray images from diagnostic 3D CT, in an effort to ease the establishment of machine learning-based approaches in fluoroscopy-guided procedures. The framework combines novel learning-based algorithms for 3D material decomposition from CT and 2D scatter estimation with fast, analytic models for energy and material dependent forward projection and noise injection. On a surrogate task, \ie the prediction of anatomical landmarks in X-ray images of the pelvis, we demonstrate that models trained on DeepDRRs generalize to clinical data without the need of re-training or domain adaptation, while the same model trained on conventional DRRs is unable to perform. Our future work will focus on improving volumetric segmentation by introducing more materials, in particular metal, and scatter estimation that could benefit from a larger training set size. 
In conclusion, we understand realistic \emph{in silico} generation of X-ray images, \eg using the proposed framework, as a catalyst designed to benefit the implementation of machine learning in fluoroscopy-guided procedures. Our framework seamlessly integrates with the software environment currently used for machine learning and will be made open-source at the time of publication.

%
%

\end{document}